\documentclass[aps,prd,onecolumn,groupedaddress,showpacs,nofootinbib,amssymb]{revtex4}
\usepackage[dvips]{graphicx}
\usepackage{amssymb}
\usepackage{amsmath}
\usepackage{graphicx,,color}
\usepackage{amsfonts}
\usepackage{bm}
\usepackage{cancel}
\usepackage{comment}

\newcommand\be{\begin{equation}}
\newcommand\ee{\end{equation}}

\allowdisplaybreaks[4]

\begin{document}

\tolerance=5000

\title{Inflationary Phenomenology of Einstein Gauss-Bonnet Gravity Compatible with GW170817}
\author{S.~D.~Odintsov,$^{1,2}$\,\thanks{odintsov@ieec.uab.es}
V.K.~Oikonomou,$^{3,4,5,6}$\,\thanks{v.k.oikonomou1979@gmail.com}}

\affiliation{$^{1)}$ ICREA, Passeig Luis Companys, 23, 08010 Barcelona, Spain\\
$^{2)}$ Institute of Space Sciences (IEEC-CSIC) C. Can Magrans
s/n,
08193 Barcelona, Spain\\
$^{3)}$ Department of Physics, Aristotle University of
Thessaloniki, Thessaloniki 54124,
Greece\\
$^{4)}$ Laboratory for Theoretical Cosmology, Tomsk State
University of Control Systems
and Radioelectronics, 634050 Tomsk, Russia (TUSUR)\\
$^{5)}$ Tomsk State Pedagogical University, 634061 Tomsk, Russia\\
$^{6)}$Theoretical Astrophysics, IAAT, University of T\"{u}bingen,
Germany }

\tolerance=5000

\begin{abstract}
In this work we shall study Einstein Gauss-Bonnet theories and we
investigate when these can have their gravitational wave speed
equal to the speed of light, which is unity in natural units, thus
becoming compatible with the striking event GW170817. We
demonstrate how this is possible and we show that if the scalar
coupling to the Gauss-Bonnet invariant is constrained to satisfy a
differential equation, the gravitational wave speed becomes equal
to one. Accordingly, we investigate the inflationary phenomenology
of the resulting restricted Einstein Gauss-Bonnet model, by
assuming that the slow-roll conditions hold true. As we
demonstrate, the compatibility with the observational data coming
from the Planck 2018 collaboration, can be achieved, even for a
power-law potential. We restricted ourselves to the study of the
power-law potential, due to the lack of analyticity, however more
realistic potentials can be used, in this case though the
calculations are not easy to be performed analytically. We also
pointed out that a string-corrected extension of the Einstein
Gauss-Bonnet model we studied, containing terms of the form $\sim
\xi(\phi) G^{ab}\partial_a\phi \partial_b \phi $ can also provide
a theory with gravity waves speed $c_T^2=1$ in natural units, if
the function $\xi(\phi)$ is appropriately constrained, however in
the absence of the Gauss-Bonnet term $\sim \xi(\phi) \mathcal{G}$
the gravity waves speed can never be $c_T^2=1$. Finally, we
discuss which extensions of the above models can provide
interesting cosmologies, since any combination of $f(R,X,\phi)$
gravities with the above string-corrected Einstein Gauss-Bonnet
models can yield $c_T^2=1$, with
$X=\frac{1}{2}\partial_{\mu}\phi\partial^{\mu}\phi $.
\end{abstract}


\maketitle

\section{Introduction}

Cosmology and astrophysics at present are in the era of great
reordering since the observational data offer incredible new
insights in the field. Recently, the striking observed event of
neutron star merging GW170817 \cite{GBM:2017lvd}, validated the
fact that the gravitational waves and electromagnetic waves have
the same propagation speed. This observation narrowed down
significantly the viable gravitational theories, since every
theory that predicts a gravitational wave speed $c_T^2$ different
than one, in natural units, is not considered as a viable
description of nature. Particularly, most of the Horndeski
theories of gravity and also all of the string-corrected
Gauss-Bonnet theories of gravity are no longer considered as
viable modified gravity theories, see Ref. \cite{Ezquiaga:2017ekz}
for a complete list of all the theories that are ruled out by
\cite{GBM:2017lvd}.

To this end, in this paper we shall consider the possibility of
reviving one class of string-inspired theories of gravity
\cite{Nojiri:2006je}, and particularly of the Einstein
Gauss-Bonnet theories of gravity. These theories can yield a
viable inflationary era, and also can describe successfully the
late-time acceleration era, for reviews see
\cite{Nojiri:2017ncd,Nojiri:2010wj,Nojiri:2006ri,Capozziello:2011et,Capozziello:2010zz,delaCruzDombriz:2012xy,Olmo:2011uz},
and also Refs.
\cite{Cognola:2006sp,Nojiri:2005vv,Nojiri:2005jg,Nojiri:2007te,Satoh:2008ck,Satoh:2007gn,Hikmawan:2015rze,Bamba:2014zoa,Yi:2018gse,Guo:2009uk,Guo:2010jr,Jiang:2013gza,Kanti:2015pda,vandeBruck:2017voa,Kanti:1998jd,Kawai:1999pw,Nozari:2017rta,Chakraborty:2018scm,Odintsov:2018zhw}
for an important stream of papers in the field. Our approach
toward reviving the Einstein Gauss-Bonnet theories of gravity will
be straightforward and focused on the speed of propagation of
gravitational waves, which we shall investigate when it is equal
to unity in natural units. As we will show, the imposed constraint
$c_T^2=1$, restricts the functional form of the coupling of the
scalar field to the Gauss-Bonnet invariant. After finding the
restricted form of the Einstein Gauss-Bonnet coupling, we shall
consider the inflationary phenomenology of the resulting theory.
Since the general case is difficult to tackle analytically, we
shall assume that the slow-roll conditions hold true, and we shall
examine the implications of this condition on the slow-roll
indices and the potential. Then by using an appropriately chosen
scalar potential, we shall examine the phenomenological viability
of the theory and we shall confront the theory with the latest
Planck 2018 observational data \cite{Akrami:2018odb}. A similar
work in the context of $f(R)$ gravity, in which differences in the
propagation phase of modified gravity models, was performed in
\cite{Nojiri:2017hai}.

This paper is organized as follows: In section II we shall review
the essential features of Einstein Gauss-Bonnet theories of
gravity and we shall specify the reason which makes the theory
invalid in view of the GW170817 event. In section III we shall
impose the slow-roll condition in the theory, and by using the
slow-roll indices we shall find the implications of the slow-roll
condition on the potential and the rest of the physical quantities
of the theory. Accordingly, we shall confront the theory with the
observational data. At the end of section III, we shall consider
several alternative forms of string-corrected theories, and we
indicate how these can become compatible with GW170817. Finally
the conclusions follow in the end of the paper.

Before we proceed, in this paper we shall assume that the
background metric is a flat Friedmann-Robertson-Walker (FRW)
spacetime, with line element,
\begin{equation}
\label{metricfrw} ds^2 = - dt^2 + a(t)^2 \sum_{i=1,2,3}
\left(dx^i\right)^2\, ,
\end{equation}
with $a(t)$ being the scale factor.

\section{Compatibility of Einstein Gauss-Bonnet Theory of Gravity with GW170817}

Let us first consider the simplest Einstein Gauss-Bonnet theory of
gravity, in which case the gravitational action in vacuum is,
\begin{equation}\label{egbaction}
\mathcal{S}=\int d^4x\sqrt{-g}\left(\frac{1}{2}f(R,\phi
,X)-\frac{1}{2}\xi (\phi )c_1\mathcal{G} \right)\, ,
\end{equation}
where $X=\frac{1}{2}\partial_{\mu}\phi\partial^{mu}\phi$, and
$\mathcal{G}=R^{abcd}R_{abcd}-4R^{ab}R_{ab}+R^2$ is the
Gauss-Bonnet invariant. Also the function $f(R,X,\phi)$ appearing
in the action (\ref{egbaction}) is chosen to be,
\begin{equation}\label{frphix}
f(R,\phi, X)=\frac{R}{\kappa^2}-2X-2 V(\phi)\, ,
\end{equation}
where $\kappa^2=\frac{1}{M_p^2}$, and $M_p$ is the four
dimensional Planck mass, while $V(\phi)$ is the scalar potential
of the canonical scalar field potential. Essentially, the scalar
theory is a canonical scalar field theory with scalar potential
$V(\phi)$. The gravitational action (\ref{egbaction}) is the
simplest case of string-corrected gravitational theory, but in a
later section we shall consider variant forms of this action, to
include higher derivative terms. The primordial perturbations of
this type of theory have been thoroughly investigated by
\cite{Noh:2001ia,Hwang:2005hb,Hwang:2002fp}, and we shall adopt
the notation of Ref. \cite{Noh:2001ia,Hwang:2005hb,Hwang:2002fp}
for convenience. For the FRW background metric, the equations of
motion of the theory are the following,
\begin{equation}\label{eqmotion1a}
\frac{3H^2}{\kappa^2}=-2X+\frac{FR-f}{2}-3H\dot{F}+12c_1H^3\dot{\xi}\,
,
\end{equation}
\begin{equation}\label{eqmotion2a}
\frac{-2\dot{H}-3H^2}{\kappa^2}=-\frac{RF-f}{2}-\frac{1}{3}\left(12c_1H^2\ddot{\xi}+2H(\dot{H}+H^2)\dot{\xi}
\right)\, ,
\end{equation}
\begin{equation}\label{eqmotion3a}
\ddot{\phi}+3H\dot{\phi}+f_{,\phi}+12c_1(\dot{H}+H^2)H^2\frac{\dot{\xi}}{\dot{\phi}}=0\,
.
\end{equation}
For the function $f(R,X,\phi) $ chosen as in Eq. (\ref{frphix}),
the gravitational equations of motion read,
\begin{equation}\label{eqmotion1}
\frac{3H^2}{\kappa^2}=\frac{\dot{\phi}^2}{2}+V(\phi)+12c_1H^3\dot{\xi}\,
,
\end{equation}
\begin{equation}\label{eqmotion2}
\frac{-2\dot{H}}{\kappa^2}=\dot{\phi^2}+12c_1H^3\dot{\xi}-4c_1\left(12c_1H^2\ddot{\xi}+2H(\dot{H}+H^2)\dot{\xi}
\right)\, ,
\end{equation}
\begin{equation}\label{eqmotion3}
\ddot{\phi}+3H\dot{\phi}+V_{,\phi}+12c_1(\dot{H}+H^2)H^2\frac{\dot{\xi}}{\dot{\phi}}=0\,
.
\end{equation}
We also introduce at this point, the $Q_i$ functions that will be
relevant in the sections to follow,
\begin{align}\label{qifunctions}
& Q_a=-4c_1\dot{\xi}H^2\, , \\ \notag & Q_b=-8c_1\dot{\xi}H\, ,
\\ \notag & Q_e=-16c_1\dot{\xi}\dot{H}\, , \\ \notag &
Q_f=8c_1\left(\ddot{\xi}-\dot{\xi}H \right)\, .
\end{align}
For the above theory, the general expression for the scalar
perturbation propagation wave speed is equal to,
\begin{equation}\label{scalarwavespeed}
c_A^2=1+\frac{\frac{\dot{F}+Q_a}{2F+Q_b}Q_e+\left(\frac{\dot{F}+Q_a}{2F+Q_b}
\right)^2Q_f}{\dot{\phi}^2+3\frac{(\dot{F}+Q_a)^2}{2F+Q_b}}\, ,
\end{equation}
where $F=\frac{\partial f}{\partial R}$. In addition, as it was
shown in Ref. \cite{Noh:2001ia,Hwang:2005hb,Hwang:2002fp}, the
gravitational wave propagation speed is equal to,
\begin{equation}\label{gravitywavespeed}
c_T^2=1-\frac{Q_f}{2F+Q_b}\, .
\end{equation}
At this point, the source of the non-viability of the model
(\ref{egbaction}) is apparent, and it is due to the fact that the
gravitational wave speed (\ref{gravitywavespeed}) is not equal to
unity. Therefore, if the function $Q_f$ is zero, then the
gravitational wave speed is equal to one. Therefore, we impose the
condition $Q_f=0$, which imposes the following condition on the
Gauss-Bonnet scalar function $\xi(\phi)$,
\begin{equation}\label{conditiona}
Q_f\sim \left(\ddot{\xi}-\dot{\xi}H \right)=0\, .
\end{equation}
Thus if the coupling $\xi(\phi)$ satisfies the differential
equation,
\begin{equation}\label{condition}
\ddot{\xi}-\dot{\xi}H =0\, ,
\end{equation}
the gravitational wave speed becomes equal to one, that is
$c_T^2=1$. The differential equation (\ref{condition}) can be
solved analytically with respect to $\dot{\xi}$, and the solution
is,
\begin{equation}\label{solution}
\dot{\xi}=\exp \left( \int_{t_i}^{t_f}H(t)dt\right)=e^N\, ,
\end{equation}
where we used the definition of the $e$-foldings number,
\begin{equation}\label{efoldingsnumberdefinition}
N=\int_{t_i}^{t_f}H(t)dt\, ,
\end{equation}
and we assumed that the integration constants are $\sim
\mathcal{O}(1)$ in reduced Planck units, for simplicity. It proves
that the explicit form of $\dot{\xi}$ is the only quantity needed
for the calculation of the slow-roll indices and of the
observational indices of inflation, so the explicit form of
$\xi(\phi)$ is redundant for our purposes. Also, by combining
equations (\ref{condition}) and (\ref{solution}), we obtain,
\begin{equation}\label{solutionddotxi}
\ddot{\xi}=H\dot{\xi}=H e^N\, ,
\end{equation}
which is also very relevant for the calculations to follow.

In conclusion, the main result of this section is Eq.
(\ref{solution}) in conjunction with (\ref{solutionddotxi}), which
when are satisfied, the gravitational wave speed of the Einstein
Gauss-Bonnet theory at hand is $c_T^2=1$ in reduced Planck units.
In the following section we shall study in detail the
phenomenological implications of the above conditions in the
Einstein Gauss-Bonnet theory at hand, when the slow-roll condition
$\dot{H}\ll H^2$ is assumed to hold true.

\section{Inflationary Phenomenology of Viable Slow-roll Einstein Gauss-Bonnet Theory of Gravity}

In this section we shall investigate the inflationary
phenomenology of the Einstein Gauss-Bonnet gravity model, with the
scalar coupling to the Gauss-Bonnet invariant satisfying Eqs.
(\ref{solution}) and (\ref{solutionddotxi}). Obviously, if the
Einstein Gauss-Bonnet satisfies Eqs. (\ref{solution}) and
(\ref{solutionddotxi}), it has a gravitational wave speed
$c_T^2=1$, thus it is compatible with the GW170817, and in this
section we shall demonstrate that the GW170817 compatible Einstein
Gauss-Bonnet model is also a viable inflationary model in the
slow-roll approximation.

The analytic calculation of the slow-roll indices and the
corresponding observational indices for inflation is quite
difficult in the general case, so we shall assume hereafter that
the slow-roll approximation holds true, which is quantified by the
following relations,
\begin{equation}\label{slowroll1}
\dot{H}\ll H^2,\,\,\,\ddot{\phi}\ll H\dot{\phi}\,
,\,\,\,V(\phi)\gg \frac{\dot{\phi}^2}{2}\, .
\end{equation}
Also the slow-roll assumption affects the slow-roll indices, and
in effect it may relate the terms involving the scalar potential,
its derivative and other functions appearing in the equations of
motion. Let us see how the gravitational equations of motion
become by taking into account the slow-roll conditions
(\ref{slowroll1}), so the last two become,
\begin{equation}\label{eqmotion2slow}
\frac{-2\dot{H}}{\kappa^2}\sim
\dot{\phi^2}+12c_1H^3\dot{\xi}-4c_1\left(12c_1H^2\ddot{\xi}+2H^3)\dot{\xi}
\right)\, ,
\end{equation}
\begin{equation}\label{eqmotion3slow}
3H\dot{\phi}+V_{,\phi}+12c_1H^4\frac{\dot{\xi}}{\dot{\phi}}\sim
0\, .
\end{equation}
By using Eq. (\ref{solutionddotxi}), and substituting
$\ddot{\xi}=H\dot{\xi}$ in Eq. (\ref{eqmotion2slow}), the latter
becomes greatly simplified, and it reads,
\begin{equation}\label{equationmotionslowsecond}
\dot{H}\simeq -\frac{1}{2}\dot{\phi}^2\kappa^2\, ,
\end{equation}
since the last two terms in Eq. (\ref{eqmotion2slow}) cancel. Thus
hereafter Eq. (\ref{equationmotionslowsecond}) will yield the
derivative of the Hubble rate $\dot{H}$. Now what is needed to
proceed is to express $\dot{\phi}$ and the Hubble rate $H$ as a
function of the scalar field $\phi$. Then by using the relation,
\begin{equation}\label{efoldingsphi}
N=\int_{\phi_k}^{\phi_f}\frac{H}{\dot{\phi}}d \phi \, ,
\end{equation}
we can express all the above quantities as a function of the
$e$-foldings number, and eventually we can confront the theory
with the observational data. Note that $\phi_k$ in Eq.
(\ref{efoldingsphi}) is the initial value of the scalar field
which is assumed to be taken at exactly the horizon crossing, and
$\phi_f$ is the value of the scalar field when inflation ends.

In order to find the implications of the slow-roll conditions on
the slow-roll indices, we must find the analytic functional form
of the slow-roll indices in terms of the scalar field. The
slow-roll indices for the theory at hand are
\cite{Noh:2001ia,Hwang:2005hb,Hwang:2002fp},
\begin{align}\label{slowrollindices}
& \epsilon_1=\frac{\dot{H}}{H^2}\,
,\,\,\,\epsilon_2=\frac{\ddot{\phi}}{H\dot{\phi}}\, , \\ \notag &
\epsilon_4=\frac{\dot{E}}{2HE}\, ,
\end{align}
where $E$ stands for,
\begin{equation}\label{epsilondef}
E=\frac{1}{\dot{\phi}^2}\left(\dot{\phi}^2+3\frac{Q_a^2}{\frac{2}{\kappa^2}+Q_b}
\right)\, ,
\end{equation}
and $Q_t=\frac{2}{\kappa^2}+\frac{1}{2}Q_b$. From the slow-roll
condition (\ref{slowroll1}) it easily obtained that
$\epsilon_2\simeq 0$, so we disregard this index hereafter. Let us
find the explicit form of the slow-roll indices $\epsilon_1$,
$\epsilon_4$, and we shall investigate the implications of the
conditions $\epsilon_1\ll 1$, $\epsilon_4\ll 1$. For the slow-roll
index $\epsilon_1$, substituting $\dot{H}$ from Eq.
(\ref{equationmotionslowsecond}), we have,
\begin{equation}\label{epsilon1analutic}
\epsilon_1=-\frac{1}{2H^2}\dot{\phi}^2\kappa^2\, ,
\end{equation}
from which it is obtained that,
\begin{equation}\label{equationslowroll1}
\kappa^2\dot{\phi}^2\ll H^2\, .
\end{equation}
Also, the function $E$ (\ref{epsilondef}), appearing in the
slow-roll index $\epsilon_4$ in Eq. (\ref{slowrollindices}), has
the following form for the theory at hand,
\begin{equation}\label{epsilonfunction}
E=\frac{48 c_1^2 H(t)^4 \dot{\xi}^2}{\kappa ^2 \dot{\phi}
\left(\frac{2}{\kappa ^2}-8 c_1 H(t)
\dot{\xi}\right)}+\frac{\dot{\phi}}{\kappa ^2}\, ,
\end{equation}
and the slow-roll index $\epsilon_4$ reads,
\begin{align}\label{epsilon4analyticform}
& \epsilon_4=\frac{96 c_1^2 H(t)^2 \Dot{H}
\dot{\xi}^2}{\left(\frac{2}{\kappa ^2}-8 c_1 H(t) \dot{\xi}\right)
\left(\frac{48 c_1^2 H(t)^4 \dot{\xi}^2}{\frac{2}{\kappa ^2}-8 c_1
H(t) \dot{\xi}}+\dot{\phi}^2\right)}+\frac{192 c_1^3 H(t)^3
\Dot{H} \dot{\xi}^3}{\left(\frac{2}{\kappa ^2}-8 c_1 H(t)
\dot{\xi}\right)^2 \left(\frac{48 c_1^2 H(t)^4
\dot{\xi}^2}{\frac{2}{\kappa ^2}-8 c_1 H(t)
\dot{\xi}}+\dot{\phi}^2\right)}\\ \notag & \frac{48 c_1^2 H(t)^3
\dot{\xi} \ddot{\xi}}{\left(\frac{2}{\kappa ^2}-8 c_1 H(t)
\dot{\xi}\right) \left(\frac{48 c_1^2 H(t)^4
\dot{\xi}^2}{\frac{2}{\kappa ^2}-8 c_1 H(t)
\dot{\xi}}+\dot{\phi}^2\right)}+\frac{192 c_1^3 H(t)^4 \dot{\xi}^2
\ddot{\xi}}{\left(\frac{2}{\kappa ^2}-8 c_1 H(t)
\dot{\xi}\right)^2 \left(\frac{48 c_1^2 H(t)^4
\dot{\xi}^2}{\frac{2}{\kappa ^2}-8 c_1 H(t)
\dot{\xi}}+\dot{\phi}^2\right)}\\ \notag & +\frac{\dot{\phi}
\ddot{\phi}}{2 H(t) \left(\frac{48 c_1^2 H(t)^4
\dot{\xi}^2}{\frac{2}{\kappa ^2}-8 c_1 H(t)
\dot{\xi}}+\dot{\phi}^2\right)}-\frac{24 c_1^2 H(t)^3 \dot{\xi}^2
\ddot{\phi}}{\dot{\phi} \left(\frac{2}{\kappa ^2}-8 c_1 H(t)
\dot{\xi}\right) \left(\frac{48 c_1^2 H(t)^4
\dot{\xi}^2}{\frac{2}{\kappa ^2}-8 c_1 H(t)
\dot{\xi}}+\dot{\phi}^2\right)}\, .
\end{align}
From the above slow-roll index we can easily understand when the
slow-roll dynamics holds true. Indeed, by assuming,
\begin{equation}\label{newslowrollconditions}
6 c_1 H(t)^3 \dot{\xi}\gg \dot{\phi}^2\, ,\,\,\,\frac{2}{\kappa
^2}\ll 8 c_1 H(t) \dot{\xi}\, ,
\end{equation}
the slow-roll index $\epsilon_4$ becomes approximately
$\epsilon_4\sim -\frac{\dot{H}}{2H^2}$, which holds true in view
of Eq. (\ref{slowroll1}). Thus the approximations
(\ref{newslowrollconditions}) are valid and we shall assume that
these complement the slow-roll conditions (\ref{slowroll1}). In
view of the condition $\frac{2}{\kappa ^2}\ll 8 c_1 H(t)
\dot{\xi}$, it holds true that $\frac{H^2}{\kappa^2}\sim
V(\phi)\gg c_1H^3\dot{\xi}$, so in view of the above and of the
slow-roll condition (\ref{slowroll1}), the equation of motion
(\ref{eqmotion1}) becomes,
\begin{equation}\label{eqmotion1final}
\frac{3H^2}{\kappa^2}\simeq V(\phi)\, ,
\end{equation}
and also Eq. (\ref{eqmotion3slow}) becomes,
\begin{equation}\label{eqmotion3slowfinal}
\dot{\phi}\simeq -12c_1H^4\frac{\dot{\xi}}{V_{,\phi}}\, .
\end{equation}
Eqs. (\ref{equationmotionslowsecond}), (\ref{eqmotion1final}) and
(\ref{eqmotion3slowfinal}), in conjunction with Eqs.
(\ref{solution}) and (\ref{solutionddotxi}), are our starting
point, since we have $\dot{H}$, $\dot{\phi}$ and the Hubble rate
$H$ expressed as functions of the scalar field $\phi$, which can
be eventually reexpressed as functions of the $e$-foldings number,
and $\dot{\xi}$, $\ddot{\xi}$ as functions of the $e$-foldings
number.

Let us calculate in detail the slow-roll indices for an
appropriately chosen potential. In general, the potential can be
arbitrarily chosen, however we shall choose a simple form in order
to provide analytic expressions for the slow-roll indices and for
the corresponding observational indices. We examined several
combinations of exponential and power-law potentials that can
yield analytic results, however the only potentials that can
provide a viable phenomenology are the power-law potentials. So
assume that the scalar potential has the form,
\begin{equation}\label{scalarpotential}
V(\phi)=V_0\phi^n\, ,
\end{equation}
where $V_0$ an arbitrary parameter of dimension sec$^{-4+n}$. In
the following we shall use Eqs. (\ref{equationmotionslowsecond}),
(\ref{eqmotion1final}) and (\ref{eqmotion3slowfinal}), in
conjunction with Eqs. (\ref{solution}) and (\ref{solutionddotxi}).
Hence combining the above, the slow-roll index $\epsilon_1$ reads,
\begin{equation}\label{slowrollindexq1ifinal}
\epsilon_1\simeq \frac{8 c_1^2 \kappa ^8 V_0 e^{2 N} \phi
^{n+2}}{3 n^2}\, ,
\end{equation}
while the slow-roll index $\epsilon_4$ reads,
\begin{equation}\label{slowrollfinalexpsilon4}
\epsilon_4\simeq \frac{9 \left(-16 \sqrt{3} c_1^3 \kappa ^{12}
e^{3 N} V_0^2 \phi ^{2 n+2}+16 c_1^2 \kappa ^6 e^{2 N} \phi ^2
\left(\kappa ^2 V_0 \phi ^n\right)^{3/2}+2 \sqrt{3} c_1 \kappa ^4
n^2 V_0 \phi ^n-3 n^2 \sqrt{\kappa ^2 V_0 \phi
^n}\right)}{\sqrt{\kappa ^2 V_0 \phi ^n} \left(4 \sqrt{3} c_1
\kappa ^2 e^N \sqrt{\kappa ^2 V_0 \phi ^n}-3\right) \left(2 \kappa
^2 \phi ^2 \left(4 \sqrt{3} c_1 \kappa ^2 e^N \sqrt{\kappa ^2 V_0
\phi ^n}-3\right)-9 n^2\right)}\, .
\end{equation}
The above need to be expressed in terms of the $e$-foldings number
$N$ defined in Eq. (\ref{efoldingsphi}), to this end we need to
determine the final value of the scalar field when inflation ends,
namely $\phi_f$. Also the slow-roll indices and the corresponding
observational indices must be evaluated at the initial value of
the scalar field at $\phi_k$, so we must solve Eq.
(\ref{efoldingsphi}) with respect to $\phi_k$, after we perform
the integration. Let us find first the final value of the scalar
field at the end of inflation, so by equating $|\epsilon_1|=1$, we
obtain,
\begin{equation}\label{finalvalueofscalarfield}
\phi_f=\left(\frac{3}{8}\right)^{\frac{1}{n+2}} \left(\frac{n^2
e^{-2 N}}{c_1^2 \kappa ^8 V_0}\right)^{\frac{1}{n+2}}\, ,
\end{equation}
so by using this and performing the integration in Eq.
(\ref{efoldingsphi}), upon inverting $N(\phi_k)$, we obtain the
function $\phi_k=\phi_k(N)$, which is,
\begin{equation}\label{phikfinal}
\phi_k=2^{-2/n} 3^{1/n} \left(c_1 \kappa ^3 \sqrt{V_0} e^Y
\left(\frac{\sqrt{3} e^{-N}
\left(\left(\frac{3}{8}\right)^{\frac{1}{n+2}} \left(\frac{n^2
e^{-2 N}}{c_1^2 \kappa ^8
V_0}\right)^{\frac{1}{n+2}}\right)^{-\frac{n}{2}}}{2 c_1 \kappa ^3
\sqrt{V_0}}-N\right)\right)^{-2/n}\, .
\end{equation}
Now we can proceed in calculating the slow-roll indices and the
corresponding observational indices of inflation. The spectral
index of the primordial scalar curvature perturbations as a
function of the slow-roll indices is equal to
\cite{Noh:2001ia,Hwang:2005hb,Hwang:2002fp},
\begin{equation}\label{spectralindexgeneral}
n_s\simeq 1+2(2\epsilon_1-\epsilon_4)\, ,
\end{equation}
which holds true when the slow-roll indices take small values. In
addition, the tensor-to-scalar ratio $r$ is equal to
\cite{Noh:2001ia,Hwang:2005hb,Hwang:2002fp},
\begin{equation}\label{tensortoscalargeneral}
r=16\Big{|}
\left(\epsilon_1-\frac{\kappa^2}{4}(-\frac{1}{H}Q_e+Q_f)
\right)\frac{1}{1+\frac{Q_b\kappa^2}{2}}c_A^3\Big{|}\, ,
\end{equation}
where we took into account that the gravitational wave speed is
equal to $c_T^2=1$ for the model at hand.

Let us now confront the theory with the observational data and
specifically with the latest Planck 2018 data which constrain the
spectral index $n_s$ and the tensor-to-scalar ratio $r$ as
follows,
\begin{equation}\label{planck2018}
n_s= 0.9649 \pm 0.0042,\,\,\,r<0.064\, .
\end{equation}
By evaluating the slow-roll index $\epsilon_1$ appearing in Eq.
(\ref{slowrollindexq1ifinal}) and $\epsilon_4$ appearing in Eq.
(\ref{slowrollfinalexpsilon4}) at $\phi=\phi_k$, with $\phi_k$
being defined as a function of the $e$-foldings number in Eq.
(\ref{phikfinal}), the resulting expressions for the observational
indices are too lengthy to present these here, however we shall
quote the values of the free parameters for which compatibility
with the observational data can be achieved. We shall work for
convenience in reduced Planck units, and the result of our
analysis is that when $c_1$ takes small values of the order
$c_1\sim \mathcal{O}(10^{-30})$ in reduced Planck units, and also
when $V_0\sim \mathcal{O}(10)$ and with $n<0$, compatibility with
the observational data can be achieved. For example when
$c_1=10^{-29.216}$, $V_0=10$ in reduced Planck units and
$n=-0.0894$, the spectral index $n_s$ and the tensor-to-scalar
ratio take the following values,
\begin{equation}\label{results}
n_s=0.96932,\,\,\,r=0.0401939\, ,
\end{equation}
which are both compatible with the observational data. Thus we
demonstrated that the Einstein Gauss-Bonnet theory with $c_T^2=1$,
and for a power-law potential can be compatible with the
observational data when the slow-roll assumption is assumed. We
need to note that the power-law potential we used, was chosen for
demonstrational reasons only, due to the fact that we wanted to
obtain analytic results. Of course, more realistic potentials can
be used in order to obtain more stringent results, however our
purpose was solely to demonstrate that the viable Einstein
Gauss-Bonnet theory that evades the GW170817 constrain on the
gravitational wave speed, can also provide a viable phenomenology.
A more detailed and thorough analysis, with more realistic scalar
potential, may require numerical analysis, so it is out of the
scope of this paper.

\subsection{Other String-corrected Theories of Gravity and
Compatibility with GW170817}

In this section we shall briefly mention several generalizations
and extensions of the Einstein Gauss-Bonnet theory we discussed in
the previous sections, that can also potentially provide viable
inflationary phenomenology and at the same time can also have the
gravitational wave speed equal to one, if some constraints are
imposed. First, a simple extension of the Einstein Gauss-Bonnet
action (\ref{egbaction}) is,
\begin{equation}\label{egbactionnew1}
\mathcal{S}=\int d^4x\sqrt{-g}\left(\frac{1}{2}f(R,\phi
,X)-\frac{1}{2}\xi (\phi )c_1\mathcal{G} \right)\, ,
\end{equation}
with the function $f(R,X,\phi)$ being an arbitrary function of its
arguments. So this case of theory may include $f(R)$ gravity with
a non-canonical scalar field, simple $k$-Essence models,
non-minimally coupled scalar theory of gravity and so on. All
these theories, yield the same gravitational wave speed as that of
Eq. (\ref{gravitywavespeed}), with $Q_f$ being the same as the one
defined in Eq. (\ref{qifunctions}). So in principle, a large class
of modified Gauss-Bonnet theories may be included. In addition,
the recently studied ghost-free Gauss-Bonnet gravities studied in
Ref. \cite{Nojiri:2019dwl}, belong in this class of models too.

More interestingly, let us consider a string-inspired corrected
action of action (\ref{egbactionnew1}), which is the following,
\begin{equation}\label{egbactionnew1}
\mathcal{S}=\int d^4x\sqrt{-g}\left(\frac{1}{2}f(R,\phi
,X)-\frac{1}{2}\xi (\phi )c_1\mathcal{G} -\frac{1}{2}\xi (\phi
)c_2G^{ab}\partial_a\phi\partial_b \phi \right)\, ,
\end{equation}
where $G^{ab}=R^{ab}-\frac{1}{2}g^{ab}R$, the Einstein tensor. In
this case, the gravitational wave speed is given by,
\begin{equation}\label{gravitywavespeednew}
c_T^2=1-\frac{Q_f}{2F+Q_b}\, ,
\end{equation}
but in this case, the function $Q_f$ is equal to,
\begin{equation}\label{newqffunction}
Q_f=8c_1\left(\ddot{\xi}-\dot{\xi}H \right)+2c_2\xi
(\phi)\dot{\phi}^2\, .
\end{equation}
So if we demand that the scalar coupling function $\xi(\phi)$ is
constrained to satisfy the following differential equation,
\begin{equation}\label{finaldifferentialequation}
8c_1\left(\ddot{\xi}-\dot{\xi}H \right)+2c_2\xi
(\phi)\dot{\phi}^2=0\, ,
\end{equation}
then the gravitational wave speed (\ref{gravitywavespeednew})
becomes $c_T^2=1$. The gravitational theory with the action
(\ref{egbactionnew1}) is the most generalized action with
string-corrections of Gauss-Bonnet type, that can be compatible
with the GW170817 results, if the coupling $\xi (\phi)$ is
restricted to satisfy the differential equation
(\ref{finaldifferentialequation}). Notice that theories containing
only the term $\sim -\frac{1}{2}\xi (\phi
)c_2G^{ab}\partial_a\phi\partial_b \phi$ can never be compatible
with GW170817, since we always obtain $c_T^2\neq 1$, irrespective
of the choice of the scalar coupling function $\xi (\phi)$. In
principle, the inflationary phenomenology of theoretical models
like that of Eq. (\ref{egbactionnew1}) can be studied in the
slow-roll approximation, however it is a much more complicated
scenario in comparison to the model (\ref{egbaction}) so we
refrain from going into details. Nevertheless, the resulting
equations of motion can be used as a reconstruction method by
specifying the Hubble rate and the function $\xi (\phi)$ which
must satisfy Eq. (\ref{finaldifferentialequation}). Then the
resulting potential that can realize such a cosmological evolution
can be found, however the calculation of the slow-roll indices
could be quite complicated.

Before closing we need to discuss an important issue related to the gravitational wave speed at cosmic times later than the inflationary era. A similar analysis to the one we performed in this paper, was carried in Ref. \cite{Gong:2017kim}, where it was also found that the coupling constant must take very small values of the order $10^{-15}$ in reduced Planck units, in order to have compatibility with the observational constraints. Thus there seems to be some sort of universal behavior in the two approaches. In our case, the gravitational wave speed of Eq. (\ref{gravitywavespeed}) has this particular form only when tensor perturbations of a flat FRW background are considered. Indeed, the perturbed metric is \cite{Hwang:2005hb},
\begin{equation}\label{greavityperturbed}
ds^2=-a^2(1+2\alpha)d\eta^2-2a^2\beta_{,\mu}d\eta dx^{\mu}+a^2\left(g_{\mu \nu}+2\varphi g_{\mu \nu}+2\gamma_{,\mu|\nu}+2C_{\mu \nu}\right)dx^{\mu}dx^{\nu}\, ,
\end{equation}
where $a d\eta= dt$, the conformal time, and the tensor perturbation is quantified mainly by $C_{\mu \nu}$, while the metric $g_{\mu \nu}$ denotes the FRW background metric. For the Einstein Gauss-Bonnet case, the differential equation that governs the evolution of the tensor gravitational perturbations is \cite{Hwang:2005hb},
\begin{equation}\label{equatonforperturbation}
\frac{1}{a^3Q_t}\frac{d}{dt}\left( a^3Q_t\dot{C}_{\mu \nu}\right)-c_T^2\frac{\Delta}{a^2}C_{\mu \nu}=0\, ,
\end{equation}
where $c_T$ is defined in Eq. (\ref{gravitywavespeed}), and $\Delta$ is the Laplacian for the FRW metric. Obviously, the above equation (\ref{equatonforperturbation}) governs the evolution of tensor perturbations before and after horizon crossing. It is thus vital to note that since $Q_f=0$ if the coupling function $\xi (\phi)$ satisfies Eqs. (\ref{conditiona}) or (\ref{condition}), the speed of the gravitational wave perturbations will always be equal to that of the speed of light, before and after horizon crossing, thus even during the matter and radiation domination era, and even at late times. However these are primordial gravitational waves, and this was exactly the focus in this work, to impose the condition $c_T^2=1$ to primordial gravitational waves, and examine the inflationary phenomenology of the model. The difference of our approach with Ref. \cite{Gong:2017kim}, is that the coupling function $\xi(\phi)$ is severely constrained to satisfy a differential equation that gives $c_T^2=1$ for the primordial tensor perturbations propagation speed.

\section{Conclusions}

In this paper we studied Einstein Gauss-Bonnet models and we
investigated when these models can be viable in view of the
striking GW170817 results which indicated that the gravitational
wave speed is $c_T^2=1$ in natural units. Specifically, the
Einstein Gauss-Bonnet models are known to have $c_T^2<1$, so in
this paper we investigated in detail when these can have
$c_T^2=1$. As we demonstrated, this can be achieved when the
scalar coupling to the Gauss-Bonnet invariant is constrained to
satisfy a differential equation. In this case, the gravitational
wave speed for the Einstein Gauss-Bonnet theory at hand becomes
equal to one. Accordingly, we assumed that the slow-roll
conditions hold true in the model at hand and we investigated the
inflationary phenomenology of the model for a specific class of
power-law scalar potentials. As we demonstrated it is possible to
achieve compatibility with the observational data, however the
results are model dependent. It is possible that a better choice
of scalar potential may yield refined inflationary phenomenology
and at the same time may also provide a successful description of
the late-time era. An example of this sort is the quintessential
inflation models
\cite{Peebles:1998qn,Dimopoulos:2017tud,Geng:2017mic,Dimopoulos:2017zvq,deHaro:2016ftq,Haro:2015ljc},
however in this case the study cannot be easily performed
analytically. We also indicated which generalized string-corrected
Gauss-Bonnet type of theories can also yield $c_T^2=1$, and thus
become compatible with the GW170817 results. As we demonstrated,
theories that also contain terms of the form $\sim \xi(\phi)
G^{ab}\partial_a\phi \partial_b \phi $, can also become compatible
with GW170817, only in the presence of an Einstein Gauss-Bonnet
coupling $\sim \xi(\phi)\mathcal{G}$, but in the absence of the
latter, these theories can never yield $c_T^2=1$. Another
important issue that we would like to discuss before closing, is
that any combination of $f(R,\phi ,X)$ gravity, in combination
with the string-corrected terms $\sim \xi(\phi)\mathcal{G}$ and
$\sim \xi(\phi) G^{ab}\partial_a\phi
\partial_b \phi $, may also provide a theory with gravity wave
speed equal to unity. This includes $f(R)$ gravity, non-minimally
coupled scalar theories and $k$-Essence theories. Also the
slow-roll condition may be replaced by the constant-roll
condition, and in this case non-Gaussianities may occur in the
power spectrum of the primordial curvature perturbations.
Specifically, a non-zero bispectrum will be obtained in the
equilateral momentum approximation. We aim to report on this last
issue in a future work.

\section*{Acknowledgments}

This work is supported by MINECO (Spain), FIS2016-76363-P, and by
project 2017 SGR247 (AGAUR, Catalonia) (S.D.O). This work is
supported by the DAAD program ``Hochschulpartnerschaften mit
Griechenland 2016'' (Projekt 57340132) (V.K.O). V.K.O is indebted
to Prof. K. Kokkotas for his hospitality in the IAAT, University
of T\"{u}bingen.

\end{document}